\documentclass[12pt]{article}
\usepackage{amsfonts}
\usepackage{bm}
\begin{document}

\renewcommand{\theequation}{\thesection.\arabic{equation}}

\title{Deformations of the Whitham
systems in the almost linear case.}

\author{A.Ya. Maltsev}

\date{
\centerline{L.D.Landau Institute for Theoretical Physics,}
\centerline{119334 ul. Kosygina 2, Moscow, maltsev@itp.ac.ru}}

\maketitle

\begin{abstract}
 We consider deformations of the Whitham systems in the case
when the initial system is close to linear one. It appears
that the almost linear case requires a special procedure
of the deformation of the Whitham system to make all the
constructions stable in the linear limit. We suggest here 
a special deformation scheme which gives a stable deformation 
of the Whitham system for an almost linear initial system.
\end{abstract}

\section{General deformation schemes.}

 We will consider deformations of the Whitham systems for
non-linear differential equations which are close to linear ones
for small values of some "non-linearity parameters"
$\lambda$.

 As is well known the Whitham method (\cite{whith1,whith2,whith3})
is connected with the slow modulations of periodic or 
quasiperiodic $m$-phase solutions of nonlinear systems

\begin{equation}
\label{insyst}
F^{i}(\bm{\varphi}, \bm{\varphi}_{t}, \bm{\varphi}_{x}, \dots )
\,\, = \,\, 0
\,\,\,\,\,\,\,\, , \,\,\,\,\, i = 1, \dots, n \,\,\, , \,\,\,
\bm{\varphi} = (\varphi^{1}, \dots, \varphi^{n})
\end{equation}
which are represented usually in the form

\begin{equation}
\label{phasesol}
\varphi^{i} (x,t) \,\, = \,\, \Phi^{i} \left( {\bf k}({\bf U})\, x
\, + \, \bm{\omega}({\bf U})\, t \, + \, \bm{\theta}_{0}, \,
{\bf U} \right)
\end{equation}
 
 In these notations the functions ${\bf k}({\bf U})$ and
$\bm{\omega}({\bf U})$ play the role of the "wave numbers"
and "frequencies" of $m$-phase solutions and $\bm{\theta}_{0}$
are the initial phase shifts. The parameters of the solutions
${\bf U} = (U^{1}, \dots, U^{N})$ can be chosen in arbitrary
way, however, we assume that they do not change under arbitrary
shifts of the initial phases $\bm{\theta}_{0}$ of solutions. 

 The functions $\Phi^{i}(\bm{\theta})$ satisfy the system

\begin{equation}
\label{phasesyst}
F^{i} \left( {\bf \Phi}, \omega^{\alpha} 
{\bf \Phi}_{\theta^{\alpha}},
k^{\beta} {\bf \Phi}_{\theta^{\beta}}, 
\dots \right) \,\,\, \equiv
\,\,\, 0
\,\,\,\,\,\,\,\, , \,\,\,\,\, i = 1, \dots, n
\end{equation}  
and we choose for every ${\bf U}$ some function
${\bf \Phi}(\bm{\theta}, {\bf U})$ as having
"zero initial phase shifts". A full set of $m$-phase solutions
of (\ref{insyst}) can then be represented in the form
(\ref{phasesol}). For $m$-phase solutions of (\ref{insyst}) 
we have then
${\bf k}({\bf U}) = (k^{1}({\bf U}), \dots, k^{m}({\bf U}))$, 
$\bm{\omega}({\bf U}) = (\omega^{1}({\bf U}), \dots,
\omega^{m}({\bf U}))$,
$\bm{\theta}_{0} = (\theta^{1}, \dots, \theta^{m})$,
where ${\bf U}= (U^{1}, \dots, U^{N})$ are parameters
of the solution. We require also that all the functions
$\Phi^{i}(\bm{\theta}, {\bf U})$ are $2\pi$-periodic with
respect to every $\theta^{\alpha}$, $\alpha = 1, \dots, m$.  

 Let us denote by $\Lambda$ the family of the functions
$\bm{\Phi}(\bm{\theta}, {\bf U})$ which depend on the parameters
${\bf U}$ in a smooth way and satisfy system (\ref{phasesyst})
for all ${\bf U}$. We will assume also that $\Lambda$ is the
maximal family having these properties.

 As it is well known the famous area of the theory of integrable 
systems connected with $m$-phase solutions was started by the 
fundamental work of S.P. Novikov (\cite{Novikov1}) where the
algebro-geometric approach to the theory of quasi-periodic solutions
was invented. An introduction of the finite-gap potentials as the
stationary points of higher KdV-equations and the investigation 
of their algebro-geometric properties has become a corner stone
of the algebro-geometric approach in the theory of solitons.
As is well known by now the theory of algebro-geometric solutions
has become much larger area then the pure theory of integrable
systems and there are many other wide areas of mathematical
physics where the varieties of Novikov potentials play a basic
role. The algebro-geometric methods have become very efficient
also in the theory of slow modulations of quasi-periodic
solutions and play in fact the fundamental role in their
consideration.

 We are going to consider here dispersion corrections
to the Whitham systems and pay the special attention to the
"almost linear" case.

 In Whitham approach the parameters
${\bf U}$ become slow functions of $x$ and $t$: 
${\bf U} = {\bf U}(X,T)$, where $X = \epsilon x$, $T = \epsilon t$
($\epsilon \rightarrow 0$).

 The functions ${\bf U}(X,T)$ should satisfy in this case
some system of differential equations (Whitham system) which
makes possible the construction of the corresponding
asymptotic solution. More precisely (see \cite{luke}), we try 
to find the asymptotic solutions
 
\begin{equation}
\label{whithsol}
\varphi^{i}(\bm{\theta}, X, T) \,\,\, = \,\,\, \sum_{k\geq0}
\Psi^{i}_{(k)} \left( {{\bf S}(X,T) \over \epsilon} +
\bm{\theta}, \, X, \, T \right) \,\, \epsilon^{k}
\end{equation}
(where all $\bm{\Psi}_{(k)}$ are $2\pi$-periodic in $\bm{\theta}$)
which satisfy the system (\ref{insyst}), i.e.
 
$$F^{i} \left( \bm{\varphi}, \epsilon \bm{\varphi}_{T},
\epsilon \bm{\varphi}_{X}, \dots \right) \,\,\, = \,\,\, 0
\,\,\,\,\,\,\,\, , \,\,\,\,\, i = 1, \dots, n $$

 The function ${\bf S}(X,T) = (S^{1}(X,T), \dots, S^{m}(X,T))$
is called a "modulated phase" of solution (\ref{whithsol}). 

 It is easy to see that the function 
$\bm{\Psi}_{(0)}(\bm{\theta}, X, T)$ should belong 
to the family of $m$-phase solutions of (\ref{insyst})
at every $X$ and $T$. We have then

\begin{equation}
\label{psi0}
\bm{\Psi}_{(0)} (\bm{\theta},X,T) \,\,\, = \,\,\,
\bm{\Phi} \left( \bm{\theta} + \bm{\theta}_{0}(X,T), {\bf U}(X,T)
\right)
\end{equation}
and

$$S^{\alpha}_{T}(X,T) \, = \, \omega^{\alpha}({\bf U}) \,\,\, ,
\,\,\,\,\, S^{\alpha}_{X}(X,T) \, = \, k^{\alpha}({\bf U}) $$
as follows from the substitution of (\ref{whithsol}) into system
(\ref{insyst}).

 The functions $\bm{\Psi}_{(k)} (\bm{\theta},X,T)$ are defined
from the linear systems

\begin{equation}
\label{ksyst}
{\hat L}^{i}_{j[{\bf U}, \bm{\theta}_{0}]}(X,T) \,\,
\Psi_{(k)}^{j} (\bm{\theta},X,T) \,\,\, = \,\,\,
f_{(k)}^{i} (\bm{\theta},X,T)
\end{equation}
where ${\hat L}^{i}_{j[{\bf U}, \bm{\theta}_{0}]}(X,T)$
is a linear operator given by the linearization of system
(\ref{phasesyst}) on solution (\ref{psi0}). The resolvability
conditions of systems (\ref{ksyst}) can be written as the
orthogonality conditions of the functions
${\bf f}_{(k)} (\bm{\theta},X,T)$ to all the
"left eigen vectors" (the eigen vectors of adjoint operator) 
$\bm{\kappa}^{(q)}_{[{\bf U}(X,T)]}
(\bm{\theta} + \bm{\theta}_{0}(X,T))$ of the operator
${\hat L}^{i}_{j[{\bf U}, \bm{\theta}_{0}]}(X,T)$ corresponding to
zero eigen-values. The resolvability conditions of (\ref{ksyst})
for $k = 1$ 

\begin{equation}
\label{1syst}
{\hat L}^{i}_{j[{\bf U}, \bm{\theta}_{0}]}(X,T) \,\,
\Psi_{(1)}^{j} (\bm{\theta},X,T) \,\,\, = \,\,\,
f_{(1)}^{i} (\bm{\theta},X,T)
\end{equation}
together with the relations
$k^{\alpha}_{T} = \omega^{\alpha}_{X} $
give the Whitham system for $m$-phase solutions of (\ref{insyst})
which plays the central role in the slow modulations approach.

 Let us say that the resolvability conditions of (\ref{ksyst})
can in fact be rather complicated in a general multi-phase
case. Indeed, we need to investigate the eigen-spaces
of the operators ${\hat L}_{[{\bf U}, \bm{\theta}_{0}]}$
and ${\hat L}^{\dagger}_{[{\bf U}, \bm{\theta}_{0}]}$
on the space of $2\pi$-periodic functions which can be rather
non-trivial in the multi-phase situation. Thus even the dimensions
of kernels of ${\hat L}_{[{\bf U}, \bm{\theta}_{0}]}$
and ${\hat L}^{\dagger}_{[{\bf U}, \bm{\theta}_{0}]}$ 
can depend in non-smooth way on the values of ${\bf U}$
so we can have a rather complicated picture on 
the ${\bf U}$-space (\cite{dm1,dm2,dobr1,dobr2}).

 These difficulties do not usually appear in the one-phase
situation ($m = 1$) where the behavior of eigen-values of 
${\hat L}_{[{\bf U}, \bm{\theta}_{0}]}$
and ${\hat L}^{\dagger}_{[{\bf U}, \bm{\theta}_{0}]}$
is usually rather regular. It is natural to introduce
some regularity conditions on the space of one-phase solutions
in this situation which will play an important role in the 
construction of asymptotic series (\ref{whithsol}). We will 
assume here that the parameters $k$ and $\omega$ can be considered
(locally) as the independent parameters on the family $\Lambda$
and the total family of solutions of (\ref{phasesyst}) depends
on $N = 2 + s$ $\,\,\,$ ($s \geq 0$) parameters $U^{\nu}$
and the initial phase $\theta_{0}$.

 Easy to see then that the functions
$\bm{\Phi}_{\theta} (\theta  +
\theta_{0}(X,T), {\bf U}(X,T))$
and \linebreak 
$\nabla_{\bm{\xi}} \, \bm{\Phi}
(\theta \, + \, \theta_{0}(X,T), {\bf U}(X,T))$
where $\bm{\xi}$ is any vector in the space of parameters
$U^{\nu}$ tangential to the surface
$k \, = \, const$, $\omega \, = \, const$
belong to the kernel of the operator
${\hat L}^{i}_{(X,T) \, j}$.

 Let us represent the space of parameters ${\bf U}$ in the form
${\bf U} = (k, \omega, {\bf n})$ where $k$
is the wave number, $\omega$ is the frequency of
one-phase solution, and ${\bf n} = (n^{1}, \dots, n^{s})$
are some additional parameters (if they exist).

\vspace{0.5cm}   

{\bf Definition 1.1.}

{\it
We call the family $\Lambda$ a full regular family of
one-phase solutions of (\ref{insyst}) if

1) The functions
$\bm{\Phi}_{\theta} (\theta, k, \omega, {\bf n})$,
$\bm{\Phi}_{n^{l}} (\theta, k, \omega, {\bf n})$
are linearly independent and give 
the full basis in the kernel of the operator
${\hat L}^{i}_{j[\theta_{0}, k, \omega,{\bf n}]}$;

2) The operator
${\hat L}^{i}_{j[\theta_{0}, k, \omega,{\bf n}]}$
has exactly $s + 1$
linearly independent "left eigen vectors"

$$\bm{\kappa}^{(q)}_{[{\bf U}]} (\theta + \theta_{0})
\,\,\, = \,\,\,
\bm{\kappa}^{(q)}_{[k, \omega, {\bf n}]}
(\theta + \theta_{0}) $$
depending on the parameters ${\bf U}$ in a smooth way and
corresponding to zero eigen-values.
}

\vspace{0.5cm}

 The Whitham system is defined in this regular situation
by the orthogonality conditions of the discrepancy
${\bf f}_{(1)}(\theta, X, T)$ to the "left eigen-vectors"
$\bm{\kappa}^{(q)}_{[k, \omega, {\bf n}]} (\theta + \theta_{0})$,
$\,\,\,$ $q = 1, \dots, s + 1$.

 As is well known, in many examples the Whitham system gives
restrictions on the parameters $U^{\nu}(X,T)$ of zero 
approximation $\bm{\Psi}_{(0)}(\theta, X, T)$ and leaves free
the parameter $\theta_{0} (X, T)$. This fact was
formulated in \cite{deform} as a general Lemma for the case
of the full regular family of ($m$-phase) solutions $\Lambda$.
In fact, it is also true in a more general ($m$-phase)
situation even without the requirements of "regularity" of the
family $\Lambda$. Let us prove here the corresponding Lemma.

\vspace{0.5cm}

{\bf Lemma 1.1.}

{\it The orthogonality conditions of all the "left eigen vectors"
of ${\hat L}^{i}_{j[{\bf U}, \bm{\theta}_{0}]}$ corresponding
to zero eigen values to the discrepancy
${\bf f}_{(1)}(\bm{\theta}, X, T)$ give restrictions on the 
functions $U^{\nu}(X,T)$ only and do not involve the functions
$\theta_{0}^{1}(X,T)$, $\dots$, $\theta_{0}^{m}(X,T)$.
}

\vspace{0.5cm}

 Proof.

 Let us represent the main term of expansion (\ref{whithsol})
in the form

$$\phi^{i}_{(0)} (\bm{\theta}, X, T) \,\, = \,\,
\Phi^{i} \left( \bm{\theta} \, + \,
{{\bf S}(X,T) + \epsilon \bm{\theta}_{0}(X,T) \over \epsilon} 
\, , \, {\bf U}(X,T) \right) $$

 Easy to see then that the part of ${\bf f}_{(1)}$ containing
the functions $\theta_{0}^{\alpha}(X,T)$ has the form

$${\tilde f}_{(1)}^{i} (\bm{\theta}, X, T) \,\, = \,\,
- \, {\partial^{\prime} F^{i} \over \partial^{\prime} \omega^{\alpha}}
\, \theta_{0T}^{\alpha} \, - \, 
{\partial^{\prime} F^{i} \over \partial^{\prime} k^{\alpha}}
\, \theta_{0X}^{\alpha} $$
where the notations 
$\partial^{\prime} F^{i} / \partial^{\prime} \omega^{\alpha}$
and $\partial^{\prime} F^{i} / \partial^{\prime} k^{\alpha}$ 
mean that we don't
consider the dependence of the functions $\Phi^{i}$ on
${\bf k}$ and $\bm{\omega}$ in (\ref{phasesyst}) and 
differentiate $F^{i}$ only with respect to the 
"explicit" $k^{\alpha}$ and $\omega^{\alpha}$ in (\ref{phasesyst})
and then put 
$\bm{\varphi} = \bm{\Phi}(\bm{\theta},{\bf k},\bm{\omega},{\bf n})$. 

 However, the full derivatives 

$${d \over d \omega^{\alpha}} \,
F^{i} \left( \bm{\Phi}, \omega^{\alpha} \bm{\Phi}_{\theta^{\alpha}},
\dots \right) \,\,\,\,\, , \,\,\,\,\,
{d \over d k^{\alpha}} \,
F^{i} \left( \bm{\Phi}, \omega^{\alpha} \bm{\Phi}_{\theta^{\alpha}},
\dots \right)$$
(including the differentiation of $\bm{\Phi}$ with respect to
$\bm{\omega}$ and ${\bf k}$) are identically zero on the family
$\Lambda$ according to (\ref{phasesyst}). So we can write that

$${\tilde f}_{(1)}^{i} \, = \,
\int_{0}^{2\pi}\!\!\!\dots\int_{0}^{2\pi} \left(
{\delta F^{i}(\bm{\theta}) \over 
\delta \Phi^{j}(\bm{\theta}^{\prime})}
\, \Phi^{j}_{\omega^{\alpha}}(\bm{\theta}^{\prime},X,T) \,
\theta_{0T}^{\alpha} \, + \, 
{\delta F^{i}(\bm{\theta}) \over 
\delta \Phi^{j}(\bm{\theta}^{\prime})}
\, \Phi^{j}_{k^{\alpha}}(\bm{\theta}^{\prime},X,T) \,
\theta_{0X}^{\alpha} \right)
{d^{m} \theta^{\prime} \over (2\pi)^{m}} $$
i.e.

$${\tilde f}_{(1)}^{i} \, = \,
{\hat L}^{i}_{j[{\bf U}, \bm{\theta}_{0}]} \,
\Phi^{j}_{\omega^{\alpha}}(\bm{\theta},X,T) 
\, \theta_{0T}^{\alpha} 
\, + \, {\hat L}^{i}_{j[{\bf U}, \bm{\theta}_{0}]} \,
\Phi^{j}_{k^{\alpha}}(\bm{\theta},X,T) 
\, \theta_{0X}^{\alpha} $$

 We have then that ${\tilde {\bf f}}_{(1)}(\theta,X,T)$
always belongs to the image of 
${\hat L}^{i}_{j[{\bf U}, \bm{\theta}_{0}]}$ without any restriction
on the functions $\theta_{0}^{\alpha} (X,T)$.

{\hfill Lemma 1.1 is proved.}

\vspace{0.5cm}

 It is easy to see also from the proof of the Lemma that the
functions $\theta_{0}^{\alpha}(X,T)$ generate the additions
$\bm{\Phi}_{\omega^{\alpha}} \, \theta_{0T}^{\alpha}$ and
$\bm{\Phi}_{k^{\alpha}} \, \theta_{0X}^{\alpha}$ to the
function $\bm{\Psi}_{(1)}(\bm{\theta}, X, T)$ which in the
main order is equivalent to the effective "renormalization"

$$\omega^{\alpha} \, \rightarrow \,
\omega^{\alpha} \, + \, \epsilon \, \theta_{0T}^{\alpha}
\,\,\,\,\, , \,\,\,\,\,
k^{\alpha} \, \rightarrow \,
k^{\alpha} \, + \, \epsilon \, \theta_{0X}^{\alpha} $$
of the parameters $(\bm{\omega}, {\bf k})$ of zero approximation
$\bm{\Phi} (\bm{\theta} + \bm{\theta}_{0}, \, {\bf U})$.

 According to the deformation procedure used in \cite{deform}
the parameters $\bm{\theta}_{0} (X,T)$ become in fact 
unnecessary after a "renormalization" of the phase
${\bf S} (X,T)$ when the unnecessary "renormalization freedom"
disappears.

 The Whitham system is a so-called system of Hydrodynamic Type,
which can be written in the form

\begin{equation}
\label{ABsyst}
A^{\nu}_{\mu} ({\bf U}) \, U^{\mu}_{T} \,\,\, = \,\,\,
B^{\nu}_{\mu} ({\bf U}) \, U^{\mu}_{X}
\end{equation}
with some matrices $A({\bf U})$ and $B({\bf U})$. In generic case
the system (\ref{ABsyst}) can be resolved w.r.t. the time   
derivatives of ${\bf U}$ and written in the evolution form

\begin{equation}
\label{HTsyst}
U^{\nu}_{T} \,\,\, = \,\,\, V^{\nu}_{\mu} ({\bf U}) \, U^{\mu}_{X}
\,\,\,\,\,\,\,\, , \,\,\,\,\, \nu = 1, \dots, N
\end{equation}
(where $V = A^{-1} B$).

 Lagrangian properties of the Whitham system were investigated
by Whitham (\cite{whith3}) who suggested also a method of
"averaging" of a Lagrangian function to get a Lagrangian
function for the Whitham system.

 Another important procedure is the procedure of "averaging"
of local Hamiltonian structures suggested by B.A. Dubrovin and
S.P. Novikov (\cite{dn1,dn2,dn3}). The Dubrovin - Novikov
procedure gives a field-theoretical Hamiltonian structure
of Hydrodynamic Type for system (\ref{HTsyst}) with a
Hamiltonian function having the hydrodynamic form
$H = \int h({\bf U}) dX$. The Dubrovin - Novikov bracket for 
system (\ref{HTsyst}) has the form

\begin{equation}
\label{DNbr}
\{U^{\nu}(X), U^{\mu}(Y)\} \,\, = \,\, g^{\nu\mu}({\bf U}) \,
\delta^{\prime}(X-Y) \, + \, b ^{\nu\mu}_{\lambda}({\bf U}) \,
U^{\lambda}_{X} \, \delta (X-Y)
\end{equation}
which is called also a local Poisson bracket of Hydrodynamic Type.

 The Hamiltonian properties of systems (\ref{HTsyst}) are
strongly correlated with their integrability properties. Thus it
was proved by S.P. Tsarev (\cite{Tsarev}) that all the
diagonalizable systems (\ref{HTsyst}) having the Dubrovin - Novikov
Hamiltonian structure can in fact be integrated (S.P. Novikov
conjecture). Actually the same is true also for the
diagonalizable systems (\ref{HTsyst}) having more general
weakly-nonlocal Mokhov-Ferapontov or Ferapontov Hamiltonian
structures. Let us say also here that the Dubrovin - Novikov
procedure of averaging of local Poisson brackets can be 
generalized also to the weakly-nonlocal case.

 The construction of asymptotic series (\ref{whithsol}) for
the case of a full regular family of (one-phase) solutions
of (\ref{insyst}) can be represented in a regular way. Namely,
provided that the Whitham system is satisfied we find the first
correction $\bm{\Psi}_{(1)}(\theta, X, T)$ at every $X$ and $T$
from system (\ref{1syst}). The function 
$\bm{\Psi}_{(1)}(\theta, X, T)$ is defined modulo the linear 
combination

\begin{equation}
\label{lincomb1}
c^{(1)} (X,T) \, \bm{\Phi}_{\theta} (\theta, X, T) \, + \,
\sum_{l=1}^{s} d^{(1)}_{l} (X,T) \, 
\bm{\Phi}_{n^{l}} (\theta, X, T)
\end{equation}
of the eigen-vectors of ${\hat L}^{i}_{j[{\bf U},\theta_{0}]}$
corresponding to zero eigen-values. The coefficients
$c^{(k)} (X,T)$ and $d^{(k)}_{l} (X,T)$ arising at every step $k$
and the initial phase $\theta_{0} (X,T)$ can be used to provide
resolvability of systems (\ref{ksyst}) in the higher orders
so we can hope to find recurrently all the corrections
$\bm{\Psi}_{(k)}(\theta, X, T)$.

 The structure of the recurrent procedure can be constructed 
in regular way. Let us note first of all that it was pointed
out by J.C. Luke that the values $c^{(1)} (X,T)$ are actually
not involved in the resolvability conditions of (\ref{ksyst})
for $k=2$ and the order $k=2$ gives restrictions on the
initial phase $\theta_{0}(X,T)$ instead (\cite{luke}, see also
\cite{dm1,maslov,BourHab,Haberman}).

 Let us say that this statement can be generalized in fact for
all the orders $k \geq 1$. Let us prove here the corresponding 
Lemma which will not actually require the "full regular family"
$\Lambda$ of solutions of (\ref{insyst}) and can be
formulated in fact in the most general ($m$-phase) situation.
We will assume just that all the corrections 
$\bm{\Psi}_{(1)}$, $\dots$, $\bm{\Psi}_{(k)}$ are found in
asymptotic solution (\ref{whithsol}) in general $m$-phase
case and we can write the general solution $\bm{\Psi}_{(k)}$
of (\ref{ksyst}) in the form

$$\bm{\Psi}_{(k)} (\bm{\theta}, X, T) \, = \,
\bm{\Psi}^{\prime}_{(k)} (\bm{\theta}, X, T) \, + \,
\sum_{\alpha=1}^{m} c^{(k)}_{\alpha} (X, T) \,
\bm{\Phi}_{\theta^{\alpha}} (\bm{\theta}, X, T) \, + \,
\sum_{l^{\prime}} d^{(k)}_{l^{\prime}} (X, T) \,
{\bf Q}_{l^{\prime}} (\bm{\theta}, X, T) $$   
where $\bm{\Psi}^{\prime}_{(k)} (\bm{\theta}, X, T)$
is normalized in some way. Here the functions
$\bm{\Phi}_{\theta^{\alpha}} (\bm{\theta}, X, T)$ always belong
to the kernel of the operator 
${\hat L}_{[{\bf U},\bm{\theta}_{0}]}$
and ${\bf Q}_{l^{\prime}} (\bm{\theta}, X, T)$ denote all the
other (linearly independent) vectors from 
$Ker \, {\hat L}_{[{\bf U},\bm{\theta}_{0}]}$. In our
notations we put here also

$$\bm{\Phi} (\bm{\theta}, X, T) \, \equiv \,
\bm{\Phi} (\bm{\theta} + \bm{\theta}_{0}(X,T), {\bf U}(X,T)) 
\, = \, \bm{\Psi}_{(0)} (\bm{\theta}, X, T) $$

\vspace{0.5cm}

{\bf Lemma 1.2.}

{\it The functions $c^{(k)}_{\alpha} (X, T)$ do not appear
in resolvability conditions of system (\ref{ksyst}) in the
order $k + 1$. }

\vspace{0.5cm}

 Proof.

 Assume first that $k \geq 2$. Let us look at the terms in
${\bf f}_{k+1} (\bm{\theta}, X, T)$ which contain the functions
$c^{(k)}_{\alpha} (X, T)$. Let us divide these terms in three
groups: ${\bf f}^{I}_{k+1}$, ${\bf f}^{II}_{k+1}$,
${\bf f}^{III}_{k+1}$ in the following way:

\noindent
I) There are terms corresponding to the "correction" of the value
$F^{i}(\bm{\varphi}, k^{\alpha} \bm{\varphi}_{\theta^{\alpha}},
\omega^{\beta} \bm{\varphi}_{\theta^{\beta}}, \dots )$
as a result of corrections of $\bm{\Psi}_{(0)}$ in the
$k$-th order. To describe these terms it is convenient to use again 
the fact that the correction 
$\epsilon^{k} c^{(k)}_{\alpha} (X, T) \, 
\bm{\Phi}_{\theta^{\alpha}} (\bm{\theta}, X, T)$
to $\bm{\Psi}_{(0)} (\bm{\theta}, X, T)$ is equivalent
(modulo the terms of order $\epsilon^{2k}$) to the correction
$\epsilon^{k+1} c^{(k)}_{\alpha} (X,T)$ to the phase 
$S^{\alpha}(X,T)$ of $\bm{\Psi}_{(0)} (\bm{\theta}, X, T)$.

 So the corresponding corrections of $k$-th and $k+1$-th
orders to $F^{i}$ will have the form:

a)

$$\int_{0}^{2\pi}\!\!\!\dots\int_{0}^{2\pi}
{\delta F^{i} (\bm{\theta}) \over \delta \varphi^{j}
(\bm{\theta}^{\prime}) } 
|_{\bm{\varphi} = \bm{\Phi}(\bm{\theta}, X, T)} \,\,\,\,\,
\epsilon^{k} \sum_{\alpha=1}^{m} c^{(k)}_{\alpha} (X, T) \,
\bm{\Phi}_{\theta^{\prime\alpha}} (\bm{\theta}^{\prime}, X, T)    
\, {d^{m} \theta^{\prime} \over (2\pi)^{m}} $$

 This correction has order $\epsilon^{k}$ but it does 
not appear in ${\bf f}_{(k)}$ since it describes in fact the
freedom of the choice of $\bm{\Psi}_{(k)}(\bm{\theta}, X, T)$
on the $k$-th step. It is equal to zero since all
$\bm{\Phi}_{\theta^{\alpha}}$ belong to the kernel of operator
${\hat L}_{[{\bf U},\bm{\theta}_{0}]}$. 

b) In the same way as in Lemma 1.1 the corrections of order
$(k+1)$ to $F^{i}$ can be represented in the form

$$\epsilon^{k+1} \sum_{\alpha=1}^{m}
{\partial^{\prime} F^{i} \over \partial^{\prime} \omega^{\alpha}}
\,\, c^{(k)}_{\alpha T} (X,T) \, + \, 
\epsilon^{k+1} \sum_{\alpha=1}^{m}
{\partial^{\prime} F^{i} \over \partial^{\prime} k^{\alpha}}
\,\, c^{(k)}_{\alpha X} (X,T) $$
where the notations $\partial^{\prime}$ mean again

$${\partial F^{i} (\bm{\varphi}, \omega^{\beta} 
\bm{\varphi}_{\theta^{\beta}}, \dots ) \over
\partial \omega^{\alpha}} 
|_{\bm{\varphi} = \bm{\Phi}(\bm{\theta}, X, T)} \,\,\,\,\, ,
\,\,\,\,\, {\partial F^{i} (\bm{\varphi}, \omega^{\beta}  
\bm{\varphi}_{\theta^{\beta}}, \dots ) \over
\partial k^{\alpha}}
|_{\bm{\varphi} = \bm{\Phi}(\bm{\theta}, X, T)} $$

 We denote a correction of this type by
$- \, \epsilon^{k+1} \, {\bf f}^{I}_{(k+1)} (\bm{\theta}, X, T)$
according to our notations.

 Again we can state that

$${\bf f}^{I}_{(k+1)} \,\, = \,\, 
\sum_{\alpha=1}^{m} c^{(k)}_{\alpha T} \,\,
{\hat L}^{i}_{j[{\bf U},\bm{\theta}_{0}]} \,\,
\Phi^{j}_{\omega^{\alpha}} \,\, + \,\,
\sum_{\alpha=1}^{m} c^{(k)}_{\alpha X} \,\,  
{\hat L}^{i}_{j[{\bf U},\bm{\theta}_{0}]} \,\,
\Phi^{j}_{k^{\alpha}} $$
in the same way as in Lemma 1.1 and we get then that
${\bf f}^{I}_{(k+1)}$ always belongs to the image of operator
${\hat L}_{[{\bf U},\bm{\theta}_{0}]}$.

 Let us consider now the other two groups of terms in 
${\bf f}_{(k+1)}$ containing $c^{(k)}_{\alpha} (X,T)$.

\noindent
II) The second group represent the
correction of \linebreak
${\bf f}_{(1)} [\bf{\Psi}_{(0)}] (\bm{\theta}, X, T)$
as a result of the corrections of $\bf{\Psi}_{(0)}$ in the
$k$-th order. The interesting part of this correction of
order $(k+1)$ has again the form

$$\epsilon^{k+1} \int_{0}^{2\pi}\!\!\!\dots\int_{0}^{2\pi}
\sum_{\alpha=1}^{m}
{\delta f^{i}_{(1)} (\bm{\theta}, X, T) \over
\delta \Psi^{j}_{(0)} (\bm{\theta}^{\prime}, X, T) } \,\,
c^{(k)}_{\alpha} (X,T) \,\, 
\bm{\Phi}_{\theta^{\prime\alpha}} (\bm{\theta}^{\prime}, X, T)
\, {d^{m} \theta^{\prime} \over (2\pi)^{m}} $$
and is equal in fact to
$\epsilon^{k+1} \sum_{\alpha=1}^{m} c^{(k)}_{\alpha} (X,T) \,\,
f^{i}_{(1)\theta^{\alpha}} (\bm{\theta}, X, T) $. So we have

$$f^{IIi}_{(k+1)} (\bm{\theta}, X, T) \, = \,
\sum_{\alpha=1}^{m} c^{(k)}_{\alpha} (X,T) \,\,
f^{i}_{(1)\theta^{\alpha}} (\bm{\theta}, X, T) $$

 This correction has the form of a shift of the phase of 
${\bf f}_{(1)}$ and belongs to ${\bf f}_{(k+1)}$.

\noindent
III) The third group is generated by the terms in 
${\bf f}_{(k+1)}$ which contain the functions 
$\bm{\Psi}_{(1)} (\bm{\theta}, X, T)$ and
$\bm{\Psi}_{(k)} (\bm{\theta}, X, T)$. Easy to see that
the interesting part of these terms can be written in the 
form

$$f^{IIIi}_{(k+1)} (\bm{\theta}, X, T) \, = \,
- \, \int_{0}^{2\pi}\!\!\!\dots\int_{0}^{2\pi}
{\delta^{2} F^{i}(\bm{\theta}) \over
\delta \varphi^{l}(\bm{\theta}^{\prime}) \,
\delta \varphi^{j}(\bm{\theta}^{\prime\prime}) }
|_{\bm{\varphi} = \bm{\Phi}(\bm{\theta}, X, T)} \, \times $$
$$\times \, \Psi^{j}_{(1)} (\bm{\theta}^{\prime\prime}, X, T) 
\, \sum_{\alpha=1}^{m} c^{(k)}_{\alpha} (X,T) \,\,
\Phi^{l}_{\theta^{\prime\alpha}} (\bm{\theta}^{\prime}, X, T) 
\, {d^{m} \theta^{\prime} \over (2\pi)^{m}}
{d^{m} \theta^{\prime\prime} \over (2\pi)^{m}} $$
i.e.

$$f^{IIIi}_{(k+1)} (\bm{\theta}, X, T) \, = \,
- \, \int_{0}^{2\pi}\!\!\!\dots\int_{0}^{2\pi}
\sum_{\alpha=1}^{m} c^{(k)}_{\alpha} (X,T) \,\,
\Phi^{l}_{\theta^{\prime\alpha}} (\bm{\theta}^{\prime}, X, T) \, 
\times $$
$$\times \, 
{\delta L^{i}_{j} (\bm{\theta}, \bm{\theta}^{\prime\prime})
\over \delta \varphi^{l}(\bm{\theta}^{\prime})}
|_{\bm{\varphi} = \bm{\Phi}(\bm{\theta}, X, T)} \,\,\,
\Psi^{j}_{(1)} (\bm{\theta}^{\prime\prime}, X, T) \,\,
{d^{m} \theta^{\prime} \over (2\pi)^{m}}
{d^{m} \theta^{\prime\prime} \over (2\pi)^{m}} $$
where the distribution 
$L^{i}_{j} (\bm{\theta}, \bm{\theta}^{\prime\prime})$
gives an "integral representation" of the operator
${\hat L}^{i}_{j[{\bf U},\bm{\theta}_{0}]}$.

 In a translationally invariant (in $\bm{\theta}$) case
it's not difficult to prove then the relation

$$f^{IIIi}_{(k+1)} (\bm{\theta}, X, T) \, = \,
- \, \int_{0}^{2\pi}\!\!\!\dots\int_{0}^{2\pi}   
\sum_{\alpha=1}^{m} c^{(k)}_{\alpha} (X,T) \,\,
{\partial \over \partial \theta^{\alpha}} \,
\left( L^{i}_{j} (\bm{\theta}, \bm{\theta}^{\prime\prime}) \,
\Psi^{j}_{(1)} (\bm{\theta}^{\prime\prime}, X, T) \right) \,
{d^{m} \theta^{\prime\prime} \over (2\pi)^{m}} \, +$$
$$+ \, \int_{0}^{2\pi}\!\!\!\dots\int_{0}^{2\pi}
\sum_{\alpha=1}^{m} c^{(k)}_{\alpha} (X,T) \,\,
L^{i}_{j} (\bm{\theta}, \bm{\theta}^{\prime\prime}) \,
\Psi^{j}_{(1)\theta^{\prime\prime\alpha}} 
(\bm{\theta}^{\prime\prime}, X, T) \,
{d^{m} \theta^{\prime\prime} \over (2\pi)^{m}} $$

 So we get that

$$f^{IIi}_{(k+1)} (\bm{\theta}, X, T) \, + \,
f^{IIIi}_{(k+1)} (\bm{\theta}, X, T) \, = \,
{\hat L}^{i}_{j} \, \sum_{\alpha=1}^{m}
c^{(k)}_{\alpha} (X,T) \,\, \Psi^{j}_{(1)\theta^{\alpha}}
(\bm{\theta}, X, T) $$
which belongs to the image of ${\hat L}^{i}_{j}$.
Thus we get the statement of the Lemma for $k \geq 2$.

 Let us consider now the case $k = 1$. Let us represent for
simplicity the solution $\bm{\Psi}_{(1)} (\bm{\theta}, X, T)$
in the form

$$\bm{\Psi}_{(1)} (\bm{\theta}, X, T) \, = \,
\bm{\Psi}^{\prime}_{(1)} (\bm{\theta}, X, T) \, + \,
\sum_{\alpha=1}^{m} c^{(1)}_{\alpha} (X,T) \,
\bm{\Phi}_{\theta^{\alpha}} (\bm{\theta}, X, T) $$
where the freedom connected with vectors 
${\bf Q}_{l^{\prime}} (\bm{\theta}, X, T)$ is included in
$\bm{\Psi}^{\prime}_{(1)} (\bm{\theta}, X, T)$.

 It is not difficult to see then that in order $\epsilon^{2}$
we will have all the terms described above for $k \geq 2$ and
one extra term:

$$f^{IVi}_{(2)} (\bm{\theta}, X, T) \, = \,
- \, \int_{0}^{2\pi}\!\!\!\dots\int_{0}^{2\pi}
{\delta^{2} F^{i}(\bm{\theta}) \over
\delta \varphi^{j}(\bm{\theta}^{\prime}) \,
\delta \varphi^{l}(\bm{\theta}^{\prime\prime}) }
|_{\bm{\varphi} = \bm{\Phi}(\bm{\theta}, X, T)} \, \times $$
$$\times \, \sum_{\alpha,\beta = 1}^{m}
c^{(1)}_{\alpha} (X,T) \, c^{(1)}_{\beta} (X,T) \, 
\Phi^{j}_{\theta^{\prime\alpha}} (\bm{\theta}^{\prime}, X, T) \,
\Phi^{l}_{\theta^{\prime\prime\beta}} 
(\bm{\theta}^{\prime\prime}, X, T) \,\,
{d^{m} \theta^{\prime} \over (2\pi)^{m}}
{d^{m} \theta^{\prime\prime} \over (2\pi)^{m}} $$
containing the functions ${\bf c}^{(1)}(X,T)$.

 Consider the expansion of the values 
$F^{i}(\bm{\Phi}, \omega^{\alpha} \bm{\Phi}_{\theta^{\alpha}},
\dots )$ under a small shift of phase 
$\bm{\theta} \rightarrow \bm{\theta} + \delta \bm{\theta}$
where $\delta \bm{\theta} = {\bf c}^{(1)} \, \delta z$,
$\delta z \rightarrow 0$. First of all we know that all the 
orders of this expansion should be equal to zero on $\Lambda$
since the values 
$F^{i}(\bm{\Phi}, \omega^{\alpha} \bm{\Phi}_{\theta^{\alpha}},
\dots )$ remain zero on $\Lambda$ after this shift.

 The second order of this expansion consists of two parts.
The first part is equal to 
$- \, f^{IVi}_{(2)} \, (\delta z)^{2}$ while the second one
is equal to

$$\int_{0}^{2\pi}\!\!\!\dots\int_{0}^{2\pi}
{\delta F^{i}(\bm{\theta}) \over   
\delta \varphi^{j}(\bm{\theta}^{\prime}) }
|_{\bm{\varphi} = \bm{\Phi}(\bm{\theta}, X, T)} \, 
\sum_{\alpha,\beta = 1}^{m}
c^{(1)}_{\alpha} (X,T) \, c^{(1)}_{\beta} (X,T) \,
(\delta z)^{2} \, 
\Phi^{j}_{\theta^{\prime\alpha}\theta^{\prime\beta}} 
(\bm{\theta}^{\prime}, X, T) \,
{d^{m} \theta^{\prime} \over (2\pi)^{m}} $$

 We get then

$$f^{IVi}_{(2)} \, = \, {\hat L}^{i}_{j[{\bf U},\bm{\theta}_{0}]}
\, \sum_{\alpha,\beta = 1}^{m}
c^{(1)}_{\alpha} (X,T) \, c^{(1)}_{\beta} (X,T) \,
\Phi^{j}_{\theta^{\alpha}\theta^{\beta}} (\bm{\theta}, X, T) $$
i.e. ${\bf f}^{IV}_{(2)}$ belongs to 
$Im \, {\hat L}_{[{\bf U},\bm{\theta}_{0}]}$. We obtain
thus the statement of the Lemma for all $k \geq 1$.

{\hfill Lemma 1.2 is proved.}

\vspace{0.5cm}

 For the case of a full regular family of (one-phase) solutions
of system (\ref{insyst}) we can see then a regular scheme of
construction of asymptotic solution (\ref{whithsol}) if the 
Whitham system (\ref{ABsyst}) is satisfied. Namely, we find the
correction $\bm{\Psi}_{(1)}(\theta, X, T)$ from the system
(\ref{1syst}) modulo linear combination (\ref{lincomb1})
and then try to find the correction
$\bm{\Psi}_{(2)}(\theta, X, T)$. The resolvability conditions
for system (\ref{ksyst}) for $k = 2$ then give us 
restrictions on the functions $\theta_{0}(X,T)$, 
$d_{l}^{(1)}(X,T)$, $l = 1, \dots, s$. Provided that the 
corresponding conditions are satisfied we find the solutions
$\bm{\Psi}_{(2)}(\theta, X, T)$ modulo a linear combination
of the same type. Now at every step $k > 2$ we will have the
restrictions on the functions $c^{(k-2)}(X,T)$,
$d_{l}^{(k-1)}(X,T)$ and obtain the solution 
$\bm{\Psi}_{(k)}(\theta, X, T)$ modulo the linear combination

\begin{equation}
\label{lincombk}
c^{(k)} (X,T) \, \bm{\Phi}_{\theta} (\theta, X, T) \, + \,
\sum_{l=1}^{s} d^{(k)}_{l} (X,T) \,
\bm{\Phi}_{n^{l}} (\theta, X, T)
\end{equation}

 So we get a regular way of the recurrent
construction of functions $\bm{\Psi}_{(k)}(\theta, X, T)$.

 The procedure described above looks quite natural. Indeed,
the functions $\theta_{0}(X,T)$, $c^{(1)}(X,T)$,
$c^{(2)}(X,T)$, $\dots$ represent in fact corrections
to the modulated phase $S(X,T)$ while the functions
$d_{l}^{(1)}(X,T)$, $d_{l}^{(2)}(X,T)$, $\dots$ 
represent corrections to the parameters 
${\bf n} = (n^{1}, \dots, n^{s})$ of the zero approximation
$\bm{\Psi}_{(0)}(\theta, X, T)$. The Whitham system 
(\ref{ABsyst}) or (\ref{HTsyst}) gives restrictions
on the functions $\omega (X,T) = S_{T}$, $k (X,T) = S_{X}$
and $n_{l} (X,T)$ as a resolvability condition of system
(\ref{1syst}) for the first correction 
$\bm{\Psi}_{(1)}(\theta, X, T)$. It is natural then that
the resolvability conditions of system (\ref{ksyst}) for
$k \geq 2$ give restrictions on the corrections
$\theta_{0}(X,T)$, $c^{(1)}(X,T)$, $c^{(2)}(X,T)$, $\dots$ and
$d_{l}^{(1)}(X,T)$, $d_{l}^{(2)}(X,T)$, $\dots$ to
$S(X,T)$ and $n^{l}(X,T)$ in a successive order.

 The Whitham solution (\ref{whithsol}) can be rewritten also
in the form:

$$\phi^{i} (\theta,X,T,\epsilon) \,\,\, = \,\,\,
\Phi^{i} \left( {S(X,T,\epsilon) \over \epsilon}
\, + \, \theta, \,\,\, S_{X}(X,T,\epsilon), \,\,\,
S_{T}(X,T,\epsilon), \,\,\, {\bf n} (X,T,\epsilon) \right)
\,\,\, + $$
\begin{equation}
\label{ser2}
+ \,\,\, \sum_{k \geq 1} \,
{\tilde \Psi}_{(k)}^{i} \left(
{S(X,T,\epsilon) \over \epsilon} \, + \, \theta, \,
X, \, T \right) \,\,\, \epsilon^{k}
\end{equation}
where we allow the regular $\epsilon$-dependence

$$S(X,T,\epsilon) \,\,\, = \,\,\,
\sum_{k\geq0} \, S_{(k)}(X,T) \,\, \epsilon^{k} 
\,\,\,\,\,\,\,\, , \,\,\,\,\,\,\,\,
n^{l} (X,T,\epsilon) \,\,\, = \,\,\,
\sum_{k\geq0} \, n_{(k)}^{l} (X,T) \,\, \epsilon^{k} $$
of the phase $S$ and the parameters
$(k, \omega, {\bf n})$ of the zero approximation
$\bm{\Psi}_{(0)} (\theta,X,T)$ such that

$$k (X,T,\epsilon) \,\,\, = \,\,\,
S_{X}(X,T,\epsilon) \,\,\,\,\,\,\,\, , \,\,\,\,\,\,\,\,
\omega (X,T,\epsilon) \,\,\, = \,\,\,
S_{T}(X,T,\epsilon) $$

 In this approach the functions 
${\tilde{\bm{\Psi}}}_{(k)} (\theta,X,T)$ can be normalized in
some way while the functions $S_{(k)}$, $n_{(k)}^{l}$ can be used
to provide the resolvability conditions of systems (\ref{ksyst}).
For instance, the normalization of 
${\tilde{\bm{\Psi}}}_{(k)} (\theta,X,T)$ used in \cite{deform}
required that the main term of (\ref{ser2}) gives 
"the best approximation" to solution (\ref{whithsol}) such that
the rest of series (\ref{ser2}) is orthogonal to the functions

$$\bm{\Phi}_{\theta} \left( {S(X,T,\epsilon) \over \epsilon}
\, + \, \theta, \,\,\, S_{X}(X,T,\epsilon), \,\,\,
S_{T}(X,T,\epsilon), \,\,\, {\bf n} (X,T,\epsilon) \right)$$
and 
$$\bm{\Phi}_{n^{l}} \left( {S(X,T,\epsilon) \over \epsilon}
\, + \, \theta, \,\,\, S_{X}(X,T,\epsilon), \,\,\,
S_{T}(X,T,\epsilon), \,\,\, {\bf n} (X,T,\epsilon) \right)$$
at every $\epsilon$.

 The functions ${\tilde{\bm{\Psi}}}_{(k)} (\theta,X,T)$
satisfy linear systems analogous to (\ref{ksyst}), i.e.

\begin{equation}
\label{novksyst}
{\hat L}^{i}_{j[S_{(0)}, {\bf n}_{(0)}]} \,\,
{\tilde \Psi}^{j}_{(k)} (\theta,X,T) \,\,\, = \,\,\,
{\tilde f}^{i}_{(k)} (\theta,X,T)
\end{equation}

 The functions ${\tilde{\bf f}}_{(k)}$ are slightly different
from ${\bf f}_{(k)}$ since a "part of $\epsilon$-dependence"
is included now in the main term of (\ref{ser2}). 

 Let us formulate here the Lemma (\cite{deform}) about the
systems on the functions $S_{(k)}(X,T)$, $n_{(k)}^{l}(X,T)$,
($k \geq 1$) arising in this approach.

\vspace{0.5cm}

{\bf Lemma 1.3.}

{\it In the case of a full regular family of (one-phase)
solutions of (\ref{insyst}) the functions
$S_{(k)}(X,T)$,  ${\bf n}_{(k)}(X,T)$
satisfy the linearized Whitham system on the functions
$S_{(0)}(X,T)$,  ${\bf n}_{(0)}(X,T)$
with an additional right-hand part depending on the functions   
$S_{(0)}$, $\dots$, $S_{(k-1)}$,
${\bf n}_{(0)}$, $\dots$, ${\bf n}_{(k-1)}$. }

\vspace{0.5cm}

 In fact, it is not difficult to show that the functions
$c^{(k-1)}(X,T)$, $d^{(k)}_{l}(X,T)$, ($k \geq 1$) 
satisfy rather similar systems in this case.

 Let us say that we can give a definition of a full regular
family of $m$-phase solutions of (\ref{insyst}) also for the
case $m > 1$ if we require the properties of Definition 1.1
for generic ${\bf k} = (k^{1}, \dots, k^{m})$ and
$\bm{\omega} = (\omega^{1}, \dots, \omega^{m})$. Let us give 
here the corresponding definition.

\vspace{0.5cm}

{\bf Definition 1.1$^{\prime}$.}

{\it
We call the family $\Lambda$ a full regular family of  
$m$-phase solutions of (\ref{insyst}) if
 
1) The functions
$\bm{\Phi}_{\theta^{\alpha}} (\bm{\theta},
{\bf k}, \bm{\omega}, {\bf n})$,
$\bm{\Phi}_{n^{l}} (\bm{\theta}, {\bf k}, \bm{\omega}, {\bf n})$
are linearly independent and give for generic ${\bf k}$ and  
$\bm{\omega}$ the full basis in the kernel of the operator
${\hat L}^{i}_{j[\bm{\theta}_{0},{\bf k},\bm{\omega},{\bf n}]}$;

2) The operator
${\hat L}^{i}_{j[\bm{\theta}_{0},{\bf k},\bm{\omega},{\bf n}]}$
has for generic ${\bf k}$ and $\bm{\omega}$ exactly $m + s$
linearly independent "left eigen vectors"

$$\bm{\kappa}^{(q)}_{[{\bf U}]} (\bm{\theta} + \bm{\theta}_{0})
\,\,\, = \,\,\,
\bm{\kappa}^{(q)}_{[{\bf k}, \bm{\omega}, {\bf n}]}
(\bm{\theta} + \bm{\theta}_{0}) $$
depending on the parameters ${\bf U}$ in a smooth way and
corresponding to zero eigen-values.
}

\vspace{0.5cm}

 All the constructions described above can be used also for
the full regular family of $m$-phase solutions if we require
that the orthogonality conditions 

\begin{equation}   
\label{ortcond1}
\int_{0}^{2\pi}\!\!\!\dots\int_{0}^{2\pi}
\kappa^{(q)}_{[{\bf U}(X,T)]\, i}
(\bm{\theta}\, + \, \bm{\theta}_{0}(X,T)) \,\,\,
f^{i}_{(k)} (\bm{\theta},X,T) \,\,\,
{d^{m} \theta \over (2\pi)^{m}} \,\,\, = \,\,\, 0
\end{equation}
give the necessary and sufficient conditions of resolvability
of systems (\ref{ksyst}). This is a serious requirement
and it can be shown (\cite{dm1,dm2,dobr1,dobr2}) that it is not 
satisfied in the general case. However, there exist examples
where this requirement is satisfied in $m$-phase situation
so the same procedure of construction of asymptotic series
(\ref{whithsol}) (or (\ref{ser2}) can be used in this case.

 Let us assume in our further considerations that we have a
full regular family of one-phase or $m$-phase solutions of
(\ref{insyst}) and the compatibility conditions of system
(\ref{ksyst}) are defined by orthogonality conditions
(\ref{ortcond1}). Let us say again, that this situation
is given by rather specific examples in a multi-phase case. 

 As far as we know dispersive corrections to the Whitham 
systems were first considered by M.Y. Ablowitz and D.J. Benney 
(\cite{AblBenny}, also \cite{Abl1}-\cite{Abl2}) where the first 
consideration of a multi-phase Whitham method was also made.
As was pointed out in \cite{AblBenny} the higher corrections 
in Whitham method satisfy more complicated equations
including "dispersive terms" and the Whitham system (\ref{ABsyst})
should in fact contain also the higher derivatives ("dispersion")   
being considered in the next orders of $\epsilon$.

 In \cite{deform,deflag} a general procedure of deformation
of the Whitham systems based on a renormalization of parameters
was suggested. The deformations of the Whitham systems appeared
in \cite{deform,deflag} have the so-called Dubrovin-Zhang form
and were considered in connection with B.A. Dubrovin problem
of deformations of Frobenius manifolds. Let us say here some
words about this deformation scheme.

 The higher corrections to Topological Quantum Field theories
require the deformations (\cite{DubrZhang1,DubrZhang2,DubrZhang3})
of the Hydrodynamic Type hierarchies (\ref{HTsyst}) having the
form

\begin{equation}
\label{defsyst}
U^{\nu}_{T} \,\,\, = \,\,\, V^{\nu}_{\mu} ({\bf U}) \, U^{\mu}_{X}
\,\, + \,\, \sum_{k\geq2} v_{(k)}^{\nu} ({\bf U}, {\bf U}_{X},
\dots, {\bf U}_{kX}) \,\, \epsilon^{k-1}
\end{equation}
where all $v_{(k)}^{\nu}$ are smooth functions polynomial in
the derivatives ${\bf U}_{X}$, $\dots$, ${\bf U}_{kX}$ and
having degree $k$ according to the following gradation rule:

1) All the functions $f({\bf U})$ have degree $0$; 

2) The derivatives $U^{\nu}_{kX}$ have degree $k$;

3) The degree of the product of two functions having certain
degrees is equal to the sum of their degrees.

 Deformation (\ref{defsyst}) of system (\ref{HTsyst}) 
implies also the deformation of the corresponding
(bi-)Hamiltonian structures (\ref{DNbr})

$$\{U^{\nu}(X), U^{\mu}(Y)\} \,\, = \,\,
\{U^{\nu}(X), U^{\mu}(Y)\}_{0} \,\, + $$
\begin{equation}
\label{defbr}
+ \,\, \sum_{k\geq2}
\epsilon^{k-1} \, \sum_{s=0}^{k}
B^{\nu\mu}_{(k)s}({\bf U}, {\bf U}_{X}, \dots, {\bf U}_{(k-s)X})
\,\, \delta^{(s)} (X - Y)
\end{equation}
where all $B^{\nu\mu}_{(k)s}$ are polynomial w.r.t.
derivatives ${\bf U}_{X}$, $\dots$, ${\bf U}_{(k-s)X}$ and
have degree $(k-s)$.

 We call deformations of form
(\ref{defsyst})-(\ref{defbr}) deformations of Dubrovin-Zhang
type. Let us say that form (\ref{defsyst})-(\ref{defbr})
is not the only possible form of deformation of the Whitham 
system. For instance, a Lorentz-invariant scheme for nonlinear
Klein-Gordon equation was considered in \cite{Lorinv}.

 However, as we will see, the deformation procedure used in
\cite{deform,deflag,Lorinv} is not very good for  
"almost linear systems" where a non-linearity is rather
small. This instability is connected with the general 
instability of the Whitham approximation in the higher
orders for the case of the small amplitude of oscillations
which was pointed out by A.C. Newell in \cite{Newell}.
In the next chapter we will suggest a deformation
scheme of the Whitham system for such "almost linear" systems
which should describe the slow modulations of periodic
(or quasiperiodic) solutions in this situation. We will use
here the deformations of Dubrovin-Zhang form (\ref{defsyst})
although other types of gradation rules are also possible as
well.

 In general a deformation of the Whitham system can be
(in new notations) described in the following way:

 We look for a solution of (\ref{insyst}) having the form

\begin{equation}
\label{newwhithsol}
\bm{\varphi}(\bm{\theta},X,T) \, = \,
\bm{\Phi} \left({\bf S}(X,T) + \bm{\theta}, {\bf S}_{X},
{\bf S}_{T}, {\bf n} \right) \, + \,
\sum_{k \geq 1} \bm{\Psi}_{(k)} 
\left({\bf S}(X,T) + \bm{\theta}, X, T \right)
\end{equation}
where the functions $\bm{\Psi}_{(k)}$ are now local functionals
of $({\bf k}, \bm{\omega}, {\bf n})$ and there derivatives
having gradation degree $k$. We omit now the parameter 
$\epsilon$ although we put first 
${\bf S} = {\bf S}(X,T,\epsilon)$, 
${\bf n} = {\bf n}(X,T,\epsilon)$. Now the higher derivatives
of $({\bf k}, \bm{\omega}, {\bf n})$ play the role of small
parameters in the expansion according to chosen gradation rule.
The Dubrovin-Zhang gradation rule implies the following simple
definitions:

 1) The functions $k^{\alpha}(X,T) \, = \, S^{\alpha}_{X}(X,T)$,
$\omega^{\alpha}(X,T) \, = \, S^{\alpha}_{T}(X,T)$, and
$n^{l}(X,T)$ have degree 0;

 2) Every differentiation with respect to $X$ adds 1 to the 
degree of a function;

3) The degree of the product of two functions having certain
degrees is equal to the sum of their degrees.

 The functions $\bm{\Psi}_{(k)}(\bm{\theta},X,T)$ are defined
from the linear systems

\begin{equation}
\label{newksyst}
{\hat L}^{i}_{[{\bf S}_{X},{\bf S}_{T},{\bf n}]j} \, 
\Psi^{j}_{(k)} (\bm{\theta}, X, T) \,\, = \,\,
f^{i}_{(k)} (\bm{\theta}, X, T)
\end{equation}
where $f^{i}_{(k)} (\bm{\theta}, X, T)$ is the discrepancy
having gradation $k$ according to the rules introduced above.

 The functions $\bm{\Psi}_{(k)}$ are uniquely normalized
by the conditions

\begin{equation}
\label{kort1}
\int_{0}^{2\pi} \!\!\! \dots \int_{0}^{2\pi}
\sum_{i=1}^{n} \, \Phi^{i}_{\theta^{\alpha}}
\left( \bm{\theta}, \,\, {\bf S}_{X}, \, {\bf S}_{T}, \,
{\bf n} \right) \,\, \Psi^{i}_{(k)}
(\bm{\theta}, X, T) \,\,\, {d^{m} \theta \over (2\pi)^{m}}
\,\,\,\,\, = \,\,\,\,\, 0
\end{equation}

\begin{equation}   
\label{kort2}
\int_{0}^{2\pi} \!\!\! \dots \int_{0}^{2\pi}
\sum_{i=1}^{n} \, \Phi^{i}_{n^{l}}
\left( \bm{\theta}, \,\,
{\bf S}_{X}, \, {\bf S}_{T}, \,
{\bf n} \right) \,\,\, \Psi^{i}_{(k)}
(\bm{\theta}, X, T) \,\,\, {d^{m} \theta \over (2\pi)^{m}}
\,\,\,\,\, = \,\,\,\,\, 0
\end{equation}
$k \geq 1$, $\,\,$ ($\alpha \, = \, 1, \dots, m$, $\,\,$
$l \, = \, 1, \dots, s$), and are local functionals of 
$({\bf k}, \bm{\omega}, {\bf n})$ and there derivatives
having gradation degree $k$.

 The "renormalized" modulated phase ${\bf S}(X,T)$ and
parameters ${\bf n}(X,T)$ satisfy now the deformed Whitham
system

\begin{equation}
\label{deform1}
S^{\alpha}_{TT} \,\,\, = \,\,\, \sum_{k\geq1} \,\,
\sigma^{\alpha}_{(k)} \, ({\bf k}, \, \bm{\omega}, \,
{\bf n}, \, {\bf k}_{X}, \, \bm{\omega}_{X}, \,
{\bf n}_{X}, \, \dots \, )
\end{equation}

\begin{equation}
\label{deform2}
n^{l}_{T} \,\,\, = \,\,\, \sum_{k\geq1} \,\,
\eta^{l}_{(k)} \, ({\bf k}, \, \bm{\omega}, \, {\bf n}, \,   
{\bf k}_{X}, \, \bm{\omega}_{X}, \, {\bf n}_{X}, \, \dots \, )
\end{equation}
where $\sigma^{\alpha}_{(k)}$, $\eta^{l}_{(k)}$ are general
polynomials in derivatives ${\bf k}_{X}$, $\bm{\omega}_{X}$,
${\bf n}_{X}$, ${\bf k}_{XX}$, $\bm{\omega}_{XX}$,
${\bf n}_{XX}$, $\dots$ (with coefficients depending on
$({\bf k},\bm{\omega},{\bf n})$) having degree $k$.

 The functions $\bm{\sigma}_{(1)}$, $\bm{\eta}_{(1)}$
coincide with the right-hand part of the Whitham system
(\ref{insyst}) and the functions 
$\bm{\sigma}_{(k)}$, $\bm{\eta}_{(k)}$ are defined by
orthogonality conditions arising on the $k$-th order of
(\ref{newksyst}). Let us remind again that we work here with
a full regular family of ($m$-phase) solutions of (\ref{insyst}).
Besides that we imply that the orthogonality conditions of
${\bf f}_{(k)}$ to all "regular" left eigen-vectors 
$\bm{\kappa}^{(q)}$ introduced in Definition 1.1$^{\prime}$ are
equivalent to the resolvability conditions of the system
(\ref{newksyst}). The deformation of the Whitham system can
be rewritten also in parameters 
$({\bf k}, \bm{\omega}, {\bf n})$ in obvious way

$$k^{\alpha}_{T} \,\,\,\,\, = \,\,\,\,\, \omega^{\alpha}_{X} $$

\begin{equation}
\label{defwhsyst}
\omega^{\alpha}_{T} \,\,\, = \,\,\, \sum_{k\geq1} \,\,
\sigma^{\alpha}_{(k)} \, ({\bf k}, \, \bm{\omega}, \,
{\bf n}, \, {\bf k}_{X}, \, \bm{\omega}_{X}, \,
{\bf n}_{X}, \, \dots \, )
\end{equation}

$$n^{l}_{T} \,\,\, = \,\,\, \sum_{k\geq1} \,\,
\eta^{l}_{(k)} \, ({\bf k}, \, \bm{\omega}, \, {\bf n}, \,
{\bf k}_{X}, \, \bm{\omega}_{X}, \, {\bf n}_{X}, \,
\dots \, ) $$

\section{Deformation procedure for almost linear systems.}
\setcounter{equation}{0}

 Let us say however that the regular procedure of deformation 
formulated above does not behave well in the case of  
"almost linear" systems (\ref{insyst}). This means that the
procedure of deformation and the corresponding asymptotic series
(\ref{newwhithsol}) do not have a good limit for systems
(\ref{insyst}) when

$$F^{i} (\lambda, \bm{\varphi}, \bm{\varphi}_{t},
\bm{\varphi}_{x}, \dots) \, = \,
F_{0}^{i} (\bm{\varphi}, \bm{\varphi}_{t},  
\bm{\varphi}_{x}, \dots) \, + \,
F_{1}^{i} (\lambda, \bm{\varphi}, \bm{\varphi}_{t},  
\bm{\varphi}_{x}, \dots) $$
where $F_{0}^{i}$ are linear in 
$(\bm{\varphi}, \bm{\varphi}_{t}, \bm{\varphi}_{x}, \dots)$
and $F_{1}^{i} (\lambda, \bm{\varphi}, \bm{\varphi}_{t},
\bm{\varphi}_{x}, \dots) \rightarrow 0$ when
$\lambda \rightarrow 0$.

 A reason for such a behavior is that the operator 
${\hat L}^{i}_{j}(\lambda)$ can now be represented in the form

$${\hat L}^{i}_{j}(\lambda) \,\, = \,\, {\hat L}^{i}_{0j}
\, + \, {\hat L}^{i}_{1j}(\lambda) $$
where the operator ${\hat L}^{i}_{0[{\bf U},\bm{\theta}_{0}]j}$
has in fact a larger number of ("left" and "right") eigen-vectors
corresponding to zero eigen-values than the operator
${\hat L}^{i}_{[{\bf U},\bm{\theta}_{0}]j}(\lambda)$
(see \cite{Newell}, Chp. 2).

 Indeed, let us consider for instance the system

\begin{equation}
\label{KG}
\varphi_{tt} \, - \, \varphi_{xx} \, + \, \varphi \, + \,
\lambda \, \varphi^{3} \,\, = \,\, 0 \,\,\,\,\, , 
\,\,\,\,\,\,\,\, \lambda \rightarrow 0
\end{equation}

 It is well known that system (\ref{KG}) has a two-parametric
family of one-phase solutions

$$\varphi \, = \, 
\Phi_{\lambda}(kx + \omega t + \theta_{0}, \mu) 
\,\,\,\,\, , \,\,\,\,\,\,\,\, (\mu = \omega^{2} - k^{2}) $$
depending on the parameters $k$ and $\omega$.

 Since the amplitude of the solution depends on $\mu$ in a
singular way near the point $\lambda = 0$ it is more convenient
to use the parameters $k$ and $A = \Phi_{max} - \Phi_{min}$
in this situation. So we will write 

$$\varphi \,\, = \,\, \Phi_{\lambda} (k x \, + \,
\omega(k,A,\lambda) \, t \, + \theta_{0}, A) $$
in the new notations. If we choose the initial phase such that 
the function $\Phi_{\lambda} (\theta, A)$ has a local maximum at 
$\theta = 0$ then we will have the additional condition
$\Phi_{\lambda} (\theta, A) = \Phi_{\lambda} (-\theta, A)$.
We assume also that 
$\Phi_{\lambda} (\theta, A) \equiv 
\Phi_{\lambda} (\theta + 2\pi, A)$ as usually and we have in the
limit $\lambda \rightarrow 0$

$$\lim_{\lambda \rightarrow 0} \,\, \Phi_{\lambda} (\theta, A)
\,\, = \,\, A \, \cos \theta $$

 The dependence $\mu (A)$ disappears in the limit
$\lambda \rightarrow 0$ and we have $\mu = 1$ for
$\lambda = 0$.

 The operator

$${\hat L}(\lambda) \, = \, (\omega^{2} - k^{2}) \,
{d^{2} \over d \theta^{2}} \, + \, 1 \, + \,
3 \, \lambda \, \Phi_{\lambda}^{2}(\theta) $$
has only one (both "left" and "right") eigen-vector
$\kappa_{\lambda}(\theta, A) = \Phi_{\lambda, \theta}
(\theta, A)$ corresponding to zero eigen-value.
We have

$$\lim_{\lambda \rightarrow 0} \,\, \Phi_{\lambda, \theta} 
(\theta, A) \,\, = \,\, - \, A \, \sin \, \theta $$
which is an eigen-vector of the operator ${\hat L}_{0}$.
However, the function $\cos \, \theta$ is also an eigen-vector
(both "left" and "right") of the operator
${\hat L}_{0} = d^{2}/d \theta^{2} + 1$ corresponding to
zero eigen-value. As a result, the operator ${\hat L}(\lambda)$
has an eigen-vector (both "left" and "right") 
$\zeta_{\lambda}(\theta, A)$ corresponding to a "small"
eigen-value $\nu(\lambda, A)$, such that
$\nu(\lambda, A) \rightarrow 0$ for $\lambda \rightarrow 0$.

 We obtain then that even if the resolvability conditions
of systems (\ref{newksyst}) are satisfied the solutions 
$\Psi_{(k)}(\theta, X, T)$ can be singular at 
$\lambda \rightarrow 0$ if the right-hand parts 
$f_{(k)}(\theta, X, T)$ are not orthogonal to the
eigen-vector $\zeta_{\lambda}(\theta, A)$. We can see then
that it's natural to require orthogonality of all
$f_{(k)}$ to both the eigen-vectors
$\kappa_{\lambda}(\theta, A)$ and 
$\zeta_{\lambda}(\theta, A)$ to make the procedure regular
in the limit $\lambda \rightarrow 0$.

 In this situation we can not require anymore the normalization
conditions (\ref{kort1}) and just require that all the 
corrections $\Psi_{(k)}$ in expansion (\ref{newwhithsol})
are regular functions at $\lambda \rightarrow 0$. Using
these requirements we can obtain now both the deformed
Whitham system for system (\ref{insyst}) and normalization
conditions for coefficients $c^{(k)}(X,T)$ arising in the
definition of the function $\Psi_{(k)}(\theta, X, T)$.

 Let us consider now the corresponding procedure in a more
general formulation. We will omit for simplicity the
additional parameters ${\bf n}$ and consider a one-phase 
situation.

 Let us assume that we have an "almost linear" system
(\ref{insyst}) and a full regular family of (one-phase)
periodic solutions

$$\varphi^{i}(x,t) \,\, = \,\, \Phi^{i}_{\lambda} (k x \, + \,
\omega(k,A,\lambda) \, t \, + \theta_{0}, A) $$

 Here we choose the parameters $(k, A)$ instead of
$(k, \omega)$ in the limit $\lambda \rightarrow 0$ where
$A$ is some parameter playing the role of amplitude. 
We will also assume that the dependence
$\omega(k,A,\lambda)$ becomes the dispersion relation
$\omega^{(0)}(k)$ for some branch of the spectrum of
linear system

\begin{equation}
\label{linsyst}
F^{i}_{0} (\bm{\varphi}, \bm{\varphi}_{t}, \bm{\varphi}_{x},
\dots ) \,\, = \,\, 0
\end{equation}
and the function $\bm{\Phi}_{\lambda} (\theta, k, A)$
becomes the corresponding solution 
$A \, \bm{\Phi}_{0} (\theta, k)$ of the system

\begin{equation}
\label{linphase}
F^{i}_{0} \left(\bm{\Phi}, \omega^{(0)}(k) \, \bm{\Phi}_{\theta},
k \, \bm{\Phi}_{\theta}, \dots \right) \,\, = \,\, 0
\end{equation}

 It is convenient to assume also that all the systems 
(\ref{insyst}), (\ref{linsyst}), (\ref{linphase}) are written
in a real form and both the dispersion relation $\omega^{(0)}(k)$
and the functions $\Phi^{i}_{0} (\theta, k)$ are real functions.
We also have that $\bm{\Phi}_{0} (\theta + 2\pi, k) \equiv 
\bm{\Phi}_{0} (\theta, k)$ and the function
$\bm{\Phi}_{0,\theta\theta} (\theta, k)$ is proportional
to $\bm{\Phi}_{0} (\theta, k)$ in this case.

 The function $\bm{\Phi}_{\lambda, \theta} (\theta, k, A)$
is a "right" eigen-vector of the operator 
${\hat L}(\lambda)$ corresponding to zero eigen-value and 
we require that there is also exactly one "left" eigen vector
$\bm{\kappa}_{[\lambda,k,A]}$ of ${\hat L}(\lambda)$ 
corresponding to zero eigen-value for all $\lambda$.
The limit of $\bm{\Phi}_{\lambda, \theta} (\theta, k, A)$
for $\lambda \rightarrow 0$ is equal to
$A \, \bm{\Phi}_{0,\theta} (\theta, k)$ and we denote
$\bm{\kappa}_{0[k]}(\theta)$ the limit of the (normalized)
vector $\bm{\kappa}_{[\lambda,k,A]}(\theta)$ for
$\lambda \rightarrow 0$. Besides that, we assume that there
exist "left" and "right" real eigen-vectors
$\bm{\zeta}_{[\lambda, k, A]}(\theta)$ and
$\bm{\xi}_{[\lambda, k, A]}(\theta)$ of the operator
${\hat L}(\lambda)$ corresponding to a "small" eigen-value
$\nu(\lambda, k, A)$ which give the additional 
"left" and "right" real eigen-vectors 
$\bm{\zeta}_{0[k]}(\theta)$ and
$\bm{\xi}_{0[k]}(\theta)$ of operator ${\hat L}_{0}$ 
corresponding to zero eigen-value. As we said already the 
existence of the vectors $\bm{\xi}_{[\lambda, k, A]}$ and
$\bm{\zeta}_{[\lambda, k, A]}$ is connected with the
existence of additional real eigen-vectors 
$\bm{\Phi}_{0,\theta\theta} (\theta, k)$
(or $\bm{\Phi}_{0} (\theta, k)$) and
$\bm{\kappa}_{0[k],\theta}(\theta)$ lying in the kernel
of the operators ${\hat L}_{0}$ and ${\hat L}^{\dagger}_{0}$
respectively.

 We assume that the branch $(0)$ of the spectrum of
${\hat L}_{0}$ is non-degenerate and

$${\partial^{\prime} {\hat L}_{0}(\omega,k) \over
\partial^{\prime} \omega} |_{\omega = \omega^{(0)}(k)}
\,\,\, \bm{\Phi}_{0}(\theta, k) \,\, \neq \,\, 0 $$
(the notation $\partial^{\prime}/\partial^{\prime} \omega$
means here that we don't keep $\omega$ and $k$ connected
by the dispersion relation and consider them as free
parameters after the substitution 
$\partial/\partial t \rightarrow \omega \,
\partial/\partial \theta$,
$\partial/\partial x \rightarrow k \,
\partial/\partial \theta$).

 We have the identity

\begin{equation}
\label{identity}
\int_{0}^{2\pi} \kappa_{0[k]i}(\theta) \,\,
{\partial^{\prime} {\hat L}^{i}_{j0}(\omega,k) \over        
\partial^{\prime} \omega} |_{\omega = \omega^{(0)}(k)}
\,\,\, {\Phi}^{j}_{0}(\theta, k) \, 
{d \theta \over 2\pi} \,\, = \,\, 0
\end{equation}

 Indeed, expression (\ref{identity}) is a limit of
the expression

$$\int_{0}^{2\pi} \kappa_{\lambda[k,A(\omega,k)]i}(\theta) \,\,
{\partial^{\prime} F^{i}(\lambda, \bm{\varphi}, 
\omega \, \bm{\varphi}_{\theta}, k \, \bm{\varphi}_{\theta},
\dots) \over \partial^{\prime} \omega } 
|_{\bm{\varphi} = \bm{\Phi}_{\lambda}(\theta, k, A(\omega, k))}
\,\, {d \theta \over 2\pi} $$
(in parameters $(\omega, k)$) at $\lambda \rightarrow 0$.
This expression according to (\ref{phasesyst}) is equal to

$$- \, 
\int_{0}^{2\pi} \kappa_{\lambda[k,A(\omega,k)]i}(\theta) \,\,
{\hat L}^{i}_{j} (\lambda, \omega, k) \,\,
{\partial \over \partial \omega} \, 
\Phi^{j}_{\lambda} \left( \theta, k, A(\omega,k) \right) \,
{d \theta \over 2\pi} $$
which is identically zero.

 However, for the vector $\bm{\zeta}_{0[k]}$ we don't have an
identity like (\ref{identity}) in generic situation, so we will
imply

\begin{equation}
\label{neq1}
\int_{0}^{2\pi} \zeta_{0[k]i}(\theta) \,\,
{\partial^{\prime} {\hat L}^{i}_{j0}(\omega,k) \over
\partial^{\prime} \omega} |_{\omega = \omega^{(0)}(k)}
\,\,\, {\Phi}^{j}_{0}(\theta, k) \,
{d \theta \over 2\pi} \,\, \neq \,\, 0
\end{equation}
and

\begin{equation}
\label{neq2}    
\int_{0}^{2\pi} \zeta_{\lambda[k,A]i}(\theta) \,\,
{\partial^{\prime} F^{i}(\lambda, \bm{\varphi},
\omega \, \bm{\varphi}_{\theta}, k \, \bm{\varphi}_{\theta},  
\dots) \over \partial^{\prime} \omega }
|_{\bm{\varphi} = \bm{\Phi}_{\lambda}(\theta, k, A)}
\,\, {d \theta \over 2\pi} \,\, = \,\, {\cal O} (1)
\,\,\,\,\, , \,\,\,\,\, \lambda \rightarrow 0
\end{equation}

 We can suggest now a recurrent procedure of construction
of asymptotic solutions (\ref{newwhithsol}) and the
deformation of the Whitham system in the almost linear case.
As we said already we are going to require the orthogonality
of the discrepancies ${\bf f}_{(k)}(\theta, X, T)$ to both
the "left" eigen-vectors 
$\bm{\kappa}_{\lambda [k,A]}(\theta + \theta_{0})$ and
$\bm{\zeta}_{\lambda [k,A]}(\theta + \theta_{0})$
at each step. We don't put normalization conditions
(\ref{kort1}) now and use the freedom in the coefficients
$c^{k}(X,T)$ and $\theta_{0}(X,T)$ to provide orthogonality
of ${\bf f}_{(k)}$ and $\bm{\zeta}_{\lambda [k,A]}$.
So we try to find a deformation of the Whitham system
in the form

\begin{equation}
\label{kasyst}
k_{T} \, = \, \left( \omega (k, A) \right)_{X}
\,\,\,\,\,\,\,\, , \,\,\,\,\,\,\,\,
A_{T} \, = \, \sum_{k\geq1} a_{k} (k, A, k_{X}, A_{X}, \dots )
\end{equation}
where all $a_{k}$ are polynomial in 
$(k_{X}, A_{X}, k_{XX}, A_{XX}, \dots )$ and have degree $k$.
Every function $a_{k} (k, A, k_{X}, A_{X}, \dots )$ is found
as previously from the orthogonality conditions of
${\bf f}_{(k)}$ and $\bm{\kappa}_{\lambda}(\theta, X, T)$
in the $k$-th order. The system

$$k_{T} \, = \, \left( \omega (k, A) \right)_{X}
\,\,\,\,\,\,\,\, , \,\,\,\,\,\,\,\,
A_{T} \, = \,  a_{1} (k, A, k_{X}, A_{X}) $$
coincides with the Whitham system for the system (\ref{insyst}).

 The functions $\theta_{0}(X,T)$ and $c^{(1)}(X,T)$,
$c^{(2)}(X,T)$, $\dots$ are defined from the orthogonality
conditions of ${\bf f}_{(1)}$ and ${\bf f}_{(2)}$,
${\bf f}_{(2)}$, $\dots$ to the "left" eigen-vector
$\bm{\zeta}_{\lambda}(\theta, X, T)$ respectively. It's
not difficult to obtain the form of the systems arising
on the function $\theta_{0}(X,T)$ and $c^{(k)}(X,T)$.
Indeed, as we saw in the proof of Lemma 1.1 the part of
${\bf f}_{(1)}$ containing the function $\theta_{0}(X,T)$
has the form

$${\tilde {\bf f}}_{(1)}(\theta, X, T) \, = \, - \,
{\partial^{\prime} F^{i} \over \partial^{\prime} \omega } \,
\theta_{0T} \, - \,
{\partial^{\prime} F^{i} \over \partial^{\prime} k } \, 
\theta_{0X} $$

 So we obtain that the equation on $\theta_{0}(X,T)$ is
the first order linear differential equation which can be
written in the form

\begin{equation}
\label{thetasyst}
Q_{\lambda} (k, A) \, \theta_{0T} \, + \,
P_{\lambda} (k, A) \, \theta_{0X} \, = \,
R_{\lambda} (k, A, k_{X}, A_{X}) 
\end{equation}
where

$$Q_{\lambda} (k, A)  =  \int_{0}^{2\pi} \!\!
\zeta_{\lambda [k,A] i} (\theta) \,
{\partial^{\prime} F^{i} \over \partial^{\prime} \omega } \,
(\theta, k, A) \, {d \theta \over 2\pi} \,\,\, , \,\,\,
P_{\lambda} (k, A)  =  \int_{0}^{2\pi} \!\!
\zeta_{\lambda [k,A] i} (\theta) \,
{\partial^{\prime} F^{i} \over \partial^{\prime} k } \,
(\theta, k, A) \, {d \theta \over 2\pi} $$

 It can be also seen from the proof of the Lemma 1.2 that
all the systems on the functions $c^{(k)}(X,T)$ can be written
in a similar form, i.e.

\begin{equation}
\label{csyst}
Q_{\lambda} (k, A) \, c^{(k)}_{0T} \, + \,
P_{\lambda} (k, A) \, c^{(k)}_{0X} \, + \,
H_{\lambda} (k, A, k_{X}, A_{X}) \, c^{(k)} 
\, = \, R^{(k)}_{\lambda} [k, A, c^{(1)}, \dots, c^{(k-1)}]
\,\,\, , \,\,\, k \geq 2 
\end{equation}
$$Q_{\lambda} (k, A) \, c^{(1)}_{0T} \, + \,
P_{\lambda} (k, A) \, c^{(1)}_{0X} \, + \,
H_{\lambda} (k, A, k_{X}, A_{X}) \, c^{(1)} \, + \,
W_{\lambda}  (k, A) \, c^{(1)} \, c^{(1)} 
\, = \, R^{(1)}_{\lambda} [k, A]$$
where $Q_{\lambda} (k, A)$ is a non-vanishing function at
$\lambda \rightarrow 0$ according to (\ref{neq2}).

 For deformed Whitham system (\ref{kasyst}) and a
corresponding solution of the Cauchy problem it is natural
to consider also the Cauchy problem for systems
(\ref{thetasyst})-(\ref{csyst}) and define the functions
$\theta_{0}(X,T)$, $c^{(k)}(X,T)$. So, finally we get a 
recurrent procedure of deformation of the Whitham system
and the construction of asymptotic solutions (\ref{newwhithsol})
which is regular at $\lambda \rightarrow 0$. Let us note that
the deformed Whitham system here is different in general case
from that described in the first chapter because of the different
normalization of the corrections $\bm{\Psi}_{(k)}(\theta, X, T)$.

 It is not difficult to understand why normalization 
conditions (\ref{kort1}) are not very good in the almost linear 
situation. Indeed, in the case close to linear one the dependence
$A(k,\omega,\lambda)$ becomes a singular function for
$\lambda \rightarrow 0$ since $k$ and $\omega$ are not 
independent in a linear case. So, the rigid fixation of phase
$S(X,T)$ by conditions (\ref{kort1}) leads to quite an
"unstable" behavior of the main term in (\ref{newwhithsol})
depending on $S_{T}$ and $S_{X}$ as on parameters. As a result,
all the higher corrections in (\ref{newwhithsol}) become
singular functions at $\lambda \rightarrow 0$. Relations
(\ref{thetasyst})-(\ref{csyst}) describe in this case the
corrections to $S(X,T)$ which should not be included
in the parameters of the main approximation to keep all the
terms stable in the limit $\lambda \rightarrow 0$.

 To avoid this difficulty arising in the Whitham approach
it was suggested by A.C. Newell not to keep a rigid dispersion
relation between $\omega$, $k$, and $A$ and correct the Whitham
equations in the almost linear case (\cite{Newell}).

 Indeed, we can see now that the most natural way is in fact 
not to put a rigid connection between the phase $S(X,T)$ and 
the parameters $(\omega, k)$ of the main approximation

$$\bm{\Psi}_{(0)} (\theta, X,T) \, = \, \bm{\Phi} \left(
\theta + \theta_{0}(X,T), k(X,T), A(k(X,T), \omega(X,T))
\right)$$ 
if we don't want to introduce the additional parameters
$\theta_{0}(X,T)$, $c^{(k)}(X,T)$.

 Indeed we can try to write the asymptotic solution of 
(\ref{insyst}) in the form

\begin{equation}
\label{finseries}
\bm{\varphi} \, = \, \bm{\Phi} 
\left(S(X,T) + \theta, k, A \right)
\, + \, \sum_{k\geq1} \bm{\Psi}_{(k)} 
\left(S(X,T) + \theta, X, T \right)
\end{equation}
where we don't require the exact relations
$S_{T} = \omega (k,A)$, $S_{X} = k$ anymore. Instead, we allow
now corrections to the phase $S(X,T)$ and put these relations
just in the main order of our asymptotic expansion. This approach
gives us the possibility to put again the normalization
conditions (\ref{kort1}) and include the corrections generated
by the functions $\theta_{0}(X,T)$, $c^{(k)}(X,T)$ to the
phase $S(X,T)$.

 The type of the corrections to the phase $S(X,T)$ depends
on a deformation type we choose. For a deformation of
type (\ref{defsyst}) it is convenient to put

\begin{equation}
\label{defdisp}
S_{X} \, = \, k \,\,\,\,\,\,\,\, , \,\,\,\,\,\,\,\,
S_{T} \, = \, \omega (k, A) \, + \, 
\sum_{k\geq1} \omega_{(k)}
\left( k, A, k_{X}, A_{X}, \dots \right)
\end{equation}
where all $\omega_{(k)}$ are polynomial in 
$(k_{X}, A_{X}, k_{XX}, A_{XX}, \dots)$ and have degree $k$.
As previously, we assume here that any function of $k$ and $A$
has degree $0$ and every differentiation with respect to $X$
adds $1$ to the degree of a function.

 Now we put again normalization conditions (\ref{kort1})
and look for both the deformation of dispersion relation
(\ref{defdisp}) and the Whitham system which has now the form

\begin{equation} 
\label{findeform}
k_{T} \, = \, {d \over dX} \left( \omega(k,A) \, + \,
\sum_{k\geq1} \omega_{(k)}
\left( k, A, k_{X}, A_{X}, \dots \right) \right)
\,\,\,\,\, , \,\,\,\,\,
A_{T} \, = \, \sum_{k\geq1} a_{(k)}
\left( k, A, k_{X}, A_{X}, \dots \right)
\end{equation}

 System (\ref{findeform}) is a closed system on the
functions $k(X,T)$, $A(X,T)$ and it's natural to consider this 
system as a full deformation of the Whitham system in the
almost linear case.

 Now the procedure of construction of asymptotic series 
(\ref{finseries}) looks similar to the previous one
and we look for the functions 
$\bm{\Psi}_{(k)}(\theta, X, T)$ as for local functionals
of $( k, A, k_{X}, A_{X}, \dots )$ polynomial in derivatives
of $(k, A)$ and having degree $k$. The functions
$\bm{\Psi}_{(k)}(\theta, X, T)$ satisfy systems
analogous to (\ref{newksyst}) and are uniquely determined
by normalization conditions (\ref{kort1}). A difference
in this approach is that we require now the orthogonality
of all the functions ${\bf f}_{(k)}(\theta, X, T)$
to both the "left" eigen-vectors 
$\bm{\kappa}_{\lambda}(\theta, X, T)$ and
$\bm{\zeta}_{\lambda}(\theta, X, T)$ of the operator
${\hat L}^{i}_{j}(\lambda)$ corresponding to zero and "small"
eigen-values. The functions 
$\omega_{(k)} (k, A, k_{X}, A_{X}, \dots )$ and
$a_{(k)} (k, A, k_{X}, A_{X}, \dots )$ are uniquely determined
by the orthogonality conditions in the $k$-th order which gives
the recurrent procedure of construction of the deformation of  
Whitham system (\ref{findeform}) and asymptotic series
(\ref{finseries}).

 Let us call system (\ref{findeform}) the deformation of the
Whitham system in an almost linear case.

 In the limit $\lambda \rightarrow 0$ system (\ref{findeform})
gives a system describing slow modulations of solutions
of a purely linear system (\ref{insyst}). All the functions
$\bm{\Phi}_{0}(\theta)$, $\bm{\kappa}_{0}(\theta)$,
$\bm{\zeta}_{0}(\theta)$, ${\bf f}_{(k)}(\theta)$
coincide in this case with their first Fourier harmonics
and in fact only two linearly independent functions of
$\theta$ $(\bm{\Phi}_{0}(\theta), \bm{\Phi}_{0,\theta}(\theta))$
arise at every $X$ and $T$ after the substitution of
$\bm{\Phi}_{0}(\theta)$ in initial system (\ref{insyst}).
The orthogonality of all ${\bf f}_{(k)}$ to both the functions
$\bm{\kappa}_{0}(\theta)$, $\bm{\zeta}_{0}(\theta)$ gives in
this case the relations ${\bf f}_{(k)} \equiv 0$ and we have
$\bm{\Psi}_{(k)} \equiv 0$ in this situation.

 All the sums in (\ref{defdisp}) and (\ref{findeform})
contain just the finite number of terms in the linear case
and the exact solution of (\ref{insyst}) is given by the relation

$$\bm{\varphi} (\theta, X, T) \, = \, A(X,T) \,
\bm{\Phi} \left( S(X,T) + \theta, k(X,T) \right) $$
where the functions $S(X,T)$, $A(X,T)$, $k(X,T)$ satisfy the
systems (\ref{defdisp})-(\ref{findeform}).

 For the $\lambda$-expansion of systems 
(\ref{defdisp})-(\ref{findeform}) the Fourier expansion of the 
functions $\bm{\Phi}_{\lambda} (\theta, k, A)$,
$\bm{\kappa}_{\lambda [k,A]} (\theta)$,
$\bm{\zeta}_{\lambda [k,A]} (\theta)$ can be used.
Indeed, the higher harmonics of functions 
$\bm{\Phi}_{\lambda}$, $\bm{\kappa}_{\lambda}$,
$\bm{\zeta}_{\lambda}$ decrease usually as some power
of $\lambda$ so we need just a finite number of Fourier
harmonics at a given order of $\lambda$. This approach can
be connected then with the well known method of including of
a non-linearity to the slowly modulated solutions of linear 
systems (see for instance \cite{whith3}). Thus, the method of
derivation of modulation equations for the parameters of
the first Fourier harmonic of the solution
(\cite{whith3}, Chp. 15-16) can be considered as a
calculation of the first $\lambda$-correction in the system
(\ref{defdisp})-(\ref{findeform}) in this case.

 Finally, let us say just some words about multi-phase
situation. As we told above, a multi-phase situation can be 
much more complicated for the construction of asymptotic solutions
(\ref{newwhithsol}) or (\ref{finseries}). So all our remarks
should be addressed to some special cases when the orthogonality
of the discrepancies ${\bf f}_{(k)} (\bm{\theta}, X, T)$ to the
"regular" left eigen-vectors
$\bm{\kappa}^{(q)}_{\lambda [{\bf k}, {\bf A}]}(\bm{\theta})$
of the operator ${\hat L}^{i}_{j}(\lambda)$ is sufficient
for the determination of the corrections
$\bm{\Psi}_{(k)} (\bm{\theta}, X, T)$.

 We will assume according to our definition that we have $m$
linearly independent "regular" left eigen-vectors 
$\bm{\kappa}^{(q)}_{\lambda [{\bf k},{\bf A}]}(\bm{\theta})$ 
of the operator ${\hat L}(\lambda)$ corresponding to zero 
eigen-values which is equal to the number of functions
$\bm{\Phi}_{\lambda,\theta^{\alpha}} (\bm{\theta}, {\bf k}, 
{\bf A})$ giving the "regular" right eigen-vectors of 
${\hat L}(\lambda)$ corresponding to zero eigen-values.
We assume also that we have $m$ parameters
${\bf A} = (A^{1}, \dots, A^{m})$ playing the role of
amplitudes (say the amplitudes of the main Fourier
harmonics $\{\cos (\theta^{\alpha} + \theta^{\alpha}_{0}) \}$
in the expansion of $\bm{\Phi}_{\lambda} (\bm{\theta})$)
and we can express the function $\omega^{\alpha}$ in the form
$\omega^{\alpha} = \omega^{\alpha}({\bf k}, {\bf A}, \lambda)$
for rather small $\lambda$. It is natural to accept then that
in the limit $\lambda \rightarrow 0$ we have

$$\bm{\Phi}_{\lambda} (\bm{\theta}, {\bf k}, {\bf A}) 
\, \rightarrow \,
\sum_{\alpha=1}^{m} A^{\alpha} \, {\bf g}^{\alpha} \,
\cos (\theta^{\alpha} + \theta^{\alpha}_{0}) $$
where 
${\bf g}^{\alpha} = (g^{\alpha}_{1}, \dots, g^{\alpha}_{n})^{t}$
are some constant vectors and the functions
$\omega^{\alpha}({\bf k}, {\bf A}, \lambda)$ become $m$
independent dispersion relations
$\omega^{\alpha} = \omega^{\alpha}(k^{\alpha})$ also independent 
of ${\bf A}$.

 According to our general approach we will assume that we have
$m$ additional "regular" left real eigen-vectors
$\bm{\zeta}^{(q)}_{\lambda [{\bf k}, {\bf A}]}(\bm{\theta})$,
$(q = 1, \dots, m)$ of ${\hat L}(\lambda)$ corresponding to
"small" eigen-values 
$\nu^{(q)}(\lambda, {\bf k}, {\bf A})$ which give $m$
additional left eigen-vectors 
$\bm{\zeta}^{(q)}_{0 [{\bf k}]}(\bm{\theta})$ of the operator
${\hat L}_{0}$ corresponding to zero eigen-values. The number
of vectors $\bm{\zeta}^{(q)}_{0 [{\bf k}]}$ corresponds in this 
case to $m$ additional right eigen-vectors
${\bf g}^{\alpha} \, \cos (\theta^{\alpha} + \theta^{\alpha}_{0})$
of the operator ${\hat L}_{0}$ corresponding to zero 
eigen-values.

 As in the one-phase situation we will require now 
orthogonality of the functions 
${\bf f}_{(k)}(\bm{\theta}, X, T)$ to both the sets of the
left eigen-vectors 
$\{ \bm{\kappa}^{(q)}_{\lambda [{\bf k}, {\bf A}]}(\bm{\theta}) \}$ 
and
$\{ \bm{\zeta}^{(q)}_{\lambda [{\bf k}, {\bf A}]}(\bm{\theta}) \}$ 
and try to find a deformation of the Whitham system and 
the dispersion relation in the form

\begin{equation}
\label{mdefdisp}
S^{\alpha}_{X} \, = \, k^{\alpha} 
\,\,\,\,\,\,\,\, , \,\,\,\,\,\,\,\,
S^{\alpha}_{T} \, = \, \omega^{\alpha}({\bf k},{\bf A}) \, + \,
\sum_{k\geq1} \omega^{\alpha}_{(k)}
\left( {\bf k},{\bf A}, {\bf k}_{X}, {\bf A}_{X}, \dots \right)
\end{equation}
\begin{equation}
\label{mfindeform}
k^{\alpha}_{T}  =  {d \over dX} 
\left( \omega^{\alpha} ({\bf k},{\bf A})  + 
\sum_{k\geq1} \omega^{\alpha}_{(k)}
\left( {\bf k},{\bf A}, {\bf k}_{X}, {\bf A}_{X}, \dots \right) 
\right)  \,\,\, , \,\,\,
A^{\alpha}_{T}  =  \sum_{k\geq1} a^{\alpha}_{(k)}
\left( {\bf k},{\bf A}, {\bf k}_{X}, {\bf A}_{X}, \dots \right)
\end{equation}
where all the functions $\omega^{\alpha}_{(k)}$, 
$a^{\alpha}_{(k)}$ satisfy the same requirements as in the 
one-phase situation.

 Using our assumptions we can try then to find the functions
$\bm{\Psi}_{(k)}$ as local functionals of 
$({\bf k}, {\bf A})$ and their derivatives and repeat all the
steps of one-phase situation. All the functions
$a^{\alpha}_{(k)}$, $\omega^{\alpha}_{(k)}$ will be uniquely
determined in this case. However, we should remind again that
this assumption is rather serious in the $m$-phase situation
and in fact is not valid in general case.

 The author is grateful to Prof. S.P. Novikov who introduced
him to the methods of Whitham theory. The author is also
grateful to Prof. B.A. Dubrovin who suggested him the problem
of deformation of the Whitham systems and to Prof. A.C. Newell
for stimulating discussions.

 The work was partially supported by the grant of President of
Russian Federation (MD-8906.2006.2) and Russian Science Support
Foundation.

\end{document}